\documentclass{aa}
\usepackage{graphicx}
\usepackage{txfonts}
\begin{document}

\title{On the accuracy of the ALI method for solving the radiative
transfer equation} \titlerunning{Accuracy of the ALI method}

\author{L. Chevallier\inst{1} \and F. Paletou \inst{2}\thanks{Present address: Observatoire Midi-Pyr\'{e}n\'{e}es, Laboratoire d'Astrophysique (UMR 5572), 14 avenue E. Belin, F-31400 Toulouse Cedex.} \and B. Rutily
\inst{1}} \institute{ Centre de Recherche Astronomique de Lyon (UMR
5574 du CNRS), Observatoire de Lyon, 9, avenue Charles Andr\'{e},
69561 Saint-Genis-Laval cedex, France\\
\email{loic.chevallier@obs.univ-lyon1.fr} \\
\email{rutily@obs.univ-lyon1.fr} \and Observatoire de la C\^{o}te
d'Azur, D\'{e}partement G. D. Cassini (UMR 6529 du CNRS), BP 4229,
06304 Nice cedex 4, France \\ \email{fpaletou@mail.ast.obs-mip.fr} }

\date{Received 25 April 2003 / Accepted 30 July 2003}

\abstract{We solve the integral equation describing the propagation of light in an isothermal
plane-parallel atmosphere of optical thickness $\tau^*$, adopting a uniform
 thermalization parameter $\epsilon$.
The solution given by the ALI method, widely used in the field of stellar atmospheres modelling,
is compared to the exact solution. 
Graphs are given that illustrate the accuracy of the ALI solution as a function of the parameters
$\epsilon$, $\tau^*$ and optical depth variable $\tau$.
 }

\maketitle

\section{Introduction}

The solution of the radiative transfer equation (RTE) is at the heart
of the stellar atmospheres modelling, since this equation has to be
solved typically thousands of times in order to construct a realistic
model.  It is thus crucial to get a clear idea of the accuracy with
which the RTE is solved, and the effect it has on the determination of
the main physical quantities of the model: populations, electron
density, temperature, etc.  In this article, we focus on the first
point checking the ALI method for solving the integral form of the RTE, since this method
is nowadays at the basis of most numerical schemes used to determine
the radiation field in stellar atmospheres. We recall that ``ALI''
means Accelerated (or Approximate) Lambda Iteration, the Lambda
operator being defined by Eqs.~(\ref{eq_lambda})-(\ref{eq_e1}) below
for the scattering law we adopt here.
The ALI code used in this paper is a combination of an accelerated iterative method (with a diagonal $\Lambda$-operator) and a formal solver based on parabolic short characteristics. Recent reviews on this approach are Paletou (2001), Hubeny (2003) and section 3 of Trujillo Bueno (2003).

The accuracy of our ALI code is tested while applied to a well-known problem
consisting of a homogeneous, isothermal slab with isotropic and
monochromatic light scattering (Sec.~\ref{sec_2}). Indeed, this
idealized problem can be solved exactly, which allows for a
direct comparison with the solution given by the ALI method. 
This problem is very simple on physical grounds but implies analytical and numerical calculations that are far from trivial.
It contains the seeds of most of the difficulties met when solving the RTE
in a thick, highly scattering medium.  It thus provides an excellent
test for numerical codes since very accurate analytical solutions
are available (Sec.~\ref{sec_3}).  After a brief description of our
ALI code, we move to the numerical tests in Sec.~\ref{sec_4}, which is
the main part of this paper.  The link with previous studies on the
subject (Trujillo Bueno \& Fabiani Bendicho 1995, Trujillo Bueno \& Manso Sainz 1999) is finally commented in Sec.~\ref{sec_5}.

\section{The standard radiative transfer problem}\label{sec_2}

This problem consists in solving the RTE in a homogeneous
plane-parallel atmosphere of optical thickness $\tau^*>0$ (possibly
infinite); light scattering is assumed to be isotropic and
monochromatic. It is furthermore supposed that the matter is 
in local thermodynamical equilibrium with uniform temperature $T$ through the atmosphere.
The thermal source function at any frequency is then $\epsilon B(T)$,
where $\epsilon$ is the (spatially invariant) photon destruction
probability per scattering and $B(T)$ the Planck function at
temperature $T$ (frequency dependence is not mentioned).
In the absence of any external source of radiation, this problem reduces
 to solving the following integral equation for the source function $S$ (Mihalas 1978):
\begin{equation}
\label{eq_s1}
S(\tau)=\epsilon B(T)+(1-\epsilon)(\Lambda S)(\tau) \, ,
\end{equation}
where the $\Lambda$-operator for isotropic and monochromatic scattering is 
\begin{equation}
\label{eq_lambda}
(\Lambda S)(\tau)= \frac{1}{2}\int_0^{\tau^*}E_1(\vert \tau-\tau^{\prime}\vert)S(\tau^{\prime}) \,\mathrm{d}\tau^{\prime} \, .
\end{equation}
Here, $E_1$ is the first exponential integral function as defined by
\begin{equation}
\label{eq_e1}
E_1(\tau)=\int_0^1\exp(-\tau/\mu)\frac{\,\mathrm{d}\mu}{\mu}\quad(\tau>0) \, .
\end{equation}
We remind the reader that Eq. (\ref{eq_s1}) models the multiple scattering of photons of frequency $\nu$
assuming that 1) the scattering is monochromatic (or coherent) if $\nu$ belongs to a continuum,
 2) the line profile is rectangular (Milne profile) if $\nu$ belongs to a spectral line (see, e.g., Ivanov 1973, p. 57).

The solution to problem (\ref{eq_s1}) is
$S(\tau)=S(\epsilon,\tau^*,\tau)B(T)$, where $S(\epsilon,\tau^*,\tau)$
satisfies the integral equation
\begin{equation}
\label{eq_s2}
S(\epsilon,\tau^*,\tau)=\epsilon+(1-\epsilon)(\Lambda S)(\epsilon,\tau^*, \tau)
\end{equation}
depending on parameters $\epsilon$ and $\tau^*$.
Note that this function is symmetrical about the $\tau$-mid-plane: $S(\epsilon,\tau^*,\tau)= S(\epsilon,\tau^*,\tau^*-\tau)$.

This equation is the integral formulation of the RTE in our model; it
specifies the {\em standard radiative transfer problem} we intend to solve
analytically (Sec.~\ref{sec_3}) and numerically (Sec.~\ref{sec_4}).

\section{Analytical solution of the standard problem}\label{sec_3}

There are many analytical methods for solving the integral equation
(\ref{eq_s2}). The classical approach, recently reviewed by Chevallier
\& Rutily (2003, hereafter Paper I), involves the basic auxiliary
functions of radiative transfer theory in plane-parallel geometry,
namely the $H$-function for a semi-infinite space, and the $X$- and
$Y$-functions for a finite slab (Chandrasekhar 1960). The
$H$-function depends on the parameter $\epsilon$ and on an angular
variable $\mu$, taken as positive hereafter. In addition the $X$- and
$Y$-functions depend  on $\tau^*$, and we have
$X(\epsilon,\tau^*,\mu)\to H(\epsilon,\mu)$ and
$Y(\epsilon,\tau^*,\mu)\to 0$ as $\tau^* \to +\infty$.

The zero-order moments of the functions $H$, $X$, and $Y$ yield the
surface values of the solution $S$ to (\ref{eq_s2}).
The moment of the $H$-function is defined and given by
\begin{equation}
\alpha_0(\epsilon)=\int_0^1H(\epsilon,\mu)\,\mathrm{d}\mu=\frac{2}{1+\sqrt{\epsilon}}
\, ,
\end{equation}
and those of the $X$- and $Y$-functions defined as
\begin{equation}
\alpha_0(\epsilon,\tau^*)=\!\int_0^1 \!X(\epsilon,\tau^*\!,\mu)\,\mathrm{d}\mu\;,\quad \beta_0(\epsilon,\tau^*)=\!\int_0^1 \!Y(\epsilon,\tau^*\!,\mu)\,\mathrm{d}\mu
\end{equation}
are related by
\begin{equation}
\left[ 1-\frac{1-\epsilon}{2}\alpha_0(\epsilon,\tau^*) \right]^2 - \left[ \frac{1-\epsilon}{2}\beta_0(\epsilon,\tau^*) \right]^2=\epsilon \, .
\end{equation}
There is no exact expression of these moments.

In a semi-infinite atmosphere, the surface value of the solution $S$ to (\ref{eq_s2}) is
\begin{equation}
S(\epsilon,0)=1-\frac{1-\epsilon}{2}\alpha_0(\epsilon)=\sqrt{\epsilon}
\end{equation}
and it is
\begin{equation}
\label{eq_s4}
S(\epsilon,\tau^*,0)=1-\frac{1-\epsilon}{2}[\alpha_0(\epsilon,\tau^*)+\beta_0(\epsilon,\tau^*)]
\end{equation}
in a finite slab.
As $S$ is symmetrical about the $\tau$-mid-plane, $S(\epsilon,\tau^*,\tau^*)= S(\epsilon,\tau^*,0)$.
These relations were first derived by Sobolev (1957,
1958).
The former result is the famous ``$\!\sqrt{\epsilon}$-law'' for
semi-infinite media. The latter one is less known; it requires a table of
moments $( \alpha_0, \,\beta_0 )$ for numerical applications.
Such tables are available in the literature: see references in Van de Hulst
(1980, p. 225-227). Very accurate surface values of the $S$-function
can also be found in Paper I.

The calculation of the function $S$ within the slab is discussed in
detail in Paper I, which contains ten-figure tables of $S(\epsilon,
\tau^*, \tau)$ for $(\epsilon, \tau^*)$ = $(0.5,2)$, $(10^{-2},20)$,
$(10^{-4},2000)$ and $(10^{-8}, 2\times 10^8)$. In a
half-space, the internal solution is known since the end of the 50's
and it can be expressed in closed-form in terms of the
$H$-function. In a finite slab, the solution involves two
non-classical auxiliary functions $\zeta_+$ and $\zeta_-$, that are
implicitly defined by Fredholm integral equations over $[0, 1]$. These
equations can be solved very accurately, so that the solution in a
finite slab is nearly as accurate as in a half-space. The accuracy is
estimated at better than $10^{-10}$ for any value of $\epsilon,
\tau^*$ and $\tau$, which means that the solution given in Paper I can
safely be used as an accuracy test of the ALI code.
\begin{figure}[tb]
\resizebox{\hsize}{!}{\includegraphics{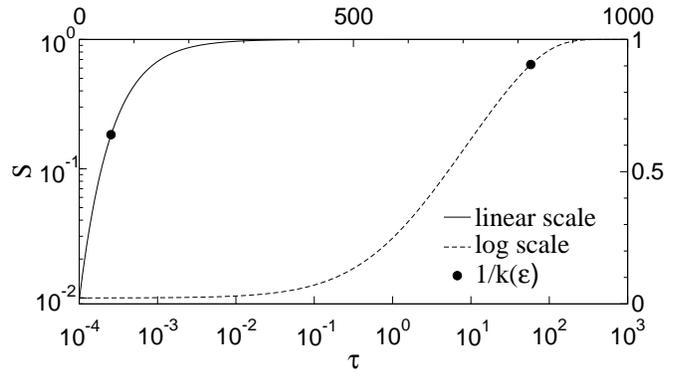}}
\caption{Source function $S(\epsilon,\tau^*,\tau)$ in a slab with $\epsilon =10^{-4}$, $\tau^*=2000$.
Right and top axes display the linear plot with infinite first derivative at $\tau=0$.
Black dots show the thermalization depth $1/k(\epsilon)$ on both scales, the definition of $k(\epsilon)$ being given in Paper I. }
\label{fig1}
\end{figure}

The general behavior of the $S$-function is shown in Fig.~\ref{fig1},
which illustrates the Table 3 of Paper I
($\epsilon = 10^{-4}$ and $\tau^* = 2000$). It can be seen that the
solution $S$ tends to 1 for large values of $\tau$ and that it drops
when $\tau$ is close to the thermalization depth $1/k(\epsilon)\approx
58$ for $\epsilon=10^{-4}$, where $k(\epsilon)$ is defined in Paper
I. It tends steeply to the surface value $S(0)$ as $\tau$ tends to 0,
{\em with an infinite derivative at 0}. It is regrettable that the
generally adopted logarithmic scale in $\tau$ obscures this essential
last point, as seen when comparing the solid and dashed curves of
Fig.~\ref{fig1}. The explanation lies in the fact that $\partial
S/\partial(\log \tau)=\tau \,\partial S/\partial \tau\to 0$ even if
$\partial S/\partial \tau\sim E_1(\tau) \to +\infty$.

\section{Comparison with ALI numerical solutions}\label{sec_4}

In this section we compare in detail the analytical
solution described in the previous section to the one given by our ALI code.
This code uses a diagonal approximate $\Lambda$-operator (Olson et al. 1986).
At each iteration, a formal solver has to be used in order to calculate the transform of the
source function by the $\Lambda$-operator. Inserting the definition
(\ref{eq_e1}) of the $E_1$-function into Eq. (\ref{eq_lambda}) and
inverting the order of integrations, the so-called formal solution to
the RTE is first calculated
\begin{eqnarray}
\label{eq_i}
\lefteqn{I(\tau ,\mu )  =  \left\lbrace \!
\begin{array}{ll}\displaystyle
 - \frac{1}{\mu } \! \int _{0}^{\tau }S(\tau^{\prime})\exp \left[ (\tau -\tau ^{\prime})/\mu \right] \,\mathrm{d}\tau ^{\prime} & \rm if\;-1\le\mu <0, \\
\displaystyle S(\tau ) & \rm if \;\mu =0, \\
\displaystyle +\frac{1}{\mu } \! \int_{\tau }^{\tau^*}S(\tau ^{\prime})\exp \left[ -(\tau ^{\prime}-\tau )/\mu \right] \,\mathrm{d}\tau ^{\prime} & \rm if\; 0<\mu \le +1. 
 \end{array} \right. }
\nonumber \\
\end{eqnarray}
Then the $\Lambda$-transform of the source function is derived, since it is here the associated mean intensity
\begin{equation}
\label{eq_ls}
(\Lambda S)(\tau )=\frac{1}{2}\int _{-1}^{+1}I(\tau ,\mu ) \,\mathrm{d}\mu \, .
\end{equation}

The formal solution (\ref{eq_i}) is calculated following the method of short characteristics
 whose basic elements can be found in Olson \& Kunasz (1987) and Kunasz \& Auer (1988).
It was further improved by the implementation of
monotonic interpolation for multi-dimensional applications (Auer \& Paletou 1994) and by Fabiani Bendicho \& Trujillo Bueno (1999) for three-dimensional
applications with horizontal periodic boundary conditions.
In the present paper, we used parabolic short characteristics.
The $\mu$-integration in (\ref{eq_ls}) is performed with the help of a Gaussian quadrature.

A numerical acceleration scheme is used so as to improve the rate
of convergence of ALI: this is the so-called Ng-acceleration
introduced in the field of radiative transfer by Auer (1987, 1991; see
also Rybicki \& Hummer 1991).

We have calculated the relative error
\begin{equation}
\label{eq_d}
d(\epsilon, \tau^*,\tau)=\left| \frac{S_\mathrm{ALI}(\epsilon,\tau^*,\tau)-S(\epsilon,\tau^*, \tau)}{S(\epsilon,\tau^*,\tau )}\right|
\end{equation}
at various optical depths, where $S(\epsilon,\tau^*,\tau)$ is the
analytical solution of Sec.~\ref{sec_3} and
$S_\mathrm{ALI}(\epsilon,\tau^*,\tau)$ is the solution given by the
ALI code.
This error corresponds to the ``true error'' defined by Auer, Fabiani Bendicho \& Trujillo Bueno (1994),
who used a finer grid to calculate $S(\epsilon, \tau^*, \tau)$.

We introduce also the maximum value of $d(\epsilon,\tau^*,\tau)$ when the
$\tau$-variable covers the domain $[0,\tau^*]$, viz.
\begin{equation}
\label{eq_dm}
d_\mathrm{M}(\epsilon,\tau^*)=\max_{0\leq\tau\leq\tau^*} d(\epsilon,\tau^*,\tau) \, .
\end{equation}

Of course $d$ and $d_\mathrm{M}$ depend on the number of iterations
$N$ performed by the ALI code during each run. Finally we define $N_\mathrm{c}$ as the
number of iterations used to reach convergence, which is the
smallest value of $N$ satisfying the condition
$|1-d_\mathrm{M}(N)/d_\mathrm{M}(+\infty)| < \varepsilon_\mathrm{c}$,
where $d_\mathrm{M}(+\infty) = d_\mathrm{M}(N=10\,000)$ and
$\varepsilon_\mathrm{c}$ is arbitrarily set to $0.01$ in the present paper.
  
The slab optical depth is discretized using a logarithmic grid, symmetric
with respect to the mid-plane, with
$n_{\tau}$ points per decade, including the $\tau=0$ point, the next point denoted by $\tau_\mathrm{m}$, and the last point $\tau=\tau^*/2$.
The angular integration in Eq. (\ref{eq_ls}) is performed with a symmetric
grid containing $n_\mu$ Gauss-Legendre points in $[0,1]$. There is no
frequency integration since light scattering has been supposed
monochromatic. In most of our calculations, we chose the values
$\tau_\mathrm{m} = 10^{-4}$, $n_\tau=9$, and $n_\mu=5$ (i.e., values
quite often adopted for stellar atmospheres modelling). Some values
of $(\epsilon,\tau^*)$ may be (0.01, 20) for a continuum, ($10^{-4},
2000$) for an ``average'' spectral line and ($10^{-8}, 2\times10^8$)
for a strong spectral line.

The quantities of interest are the maximum relative error
$d_\mathrm{M}$ and the number of iterations to reach
convergence $N_\mathrm{c}$, which depend on $\epsilon$, $\tau^*$ and numerical parameters
$\tau_\mathrm{m}$, $n_\tau$, $n_\mu$ and $N$. We first study the
variation of $d(\epsilon,\tau^*,\tau)$ with $\tau$. Then, we study the
influence of $\epsilon$, $\tau^*$, $n_\tau$ on $d_\mathrm{M}$ and
$N_\mathrm{c}$, for given $\tau_\mathrm{m}$, $n_\mu$ and for $N=N_\mathrm{c}$.

\subsection{The influence of $\tau$ and $N$}

\begin{figure}
\resizebox{\hsize}{!}{\includegraphics{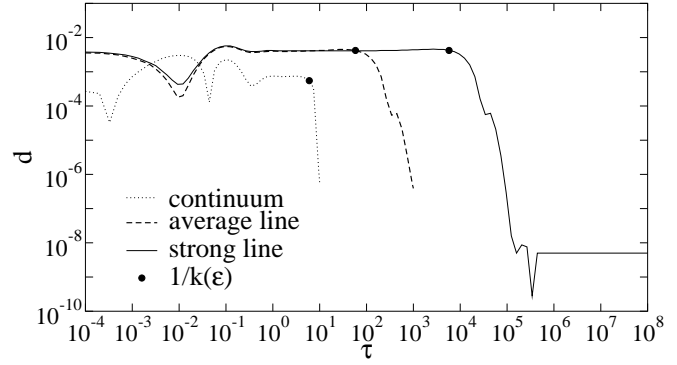}}
\caption{Relative error $d(\epsilon,\tau^*,\tau)$ of the ALI code for
$n_\tau = 9$, $n_\mu = 5$, $\tau_\mathrm{m} = 10^{-4}$ and
$N=1000$. Black dots show the thermalization depth $1/k(\epsilon)$ for
each case.}
\label{fig2}
\end{figure}

Figure~\ref{fig2} shows the variation of the relative error $\tau \to
d(\epsilon,\tau^*,\tau)$ for the three selected values of $\epsilon$
and $\tau^*$.
It can be seen that the accuracy (i.e. maximum relative error) of our ALI code is about $5\times 10^{-3}$ for the three cases studied here.
Relative error is close to this accuracy when $\tau$ is smaller than the thermalization depth $1/k(\epsilon)$ of the atmosphere (black dots on the curves), and significantly improves beyond (up to $10^{-8}$).
In photon mean free path units, the thermalization depth is 6, 58 and 5774 for $\epsilon = 10^{-2}, 10^{-4}$ and $10^{-8}$ respectively.
Note that the surface relative error is a good estimator of the accuracy in spectral lines, but not in the continuum.
\begin{figure}
\resizebox{\hsize}{!}{\includegraphics{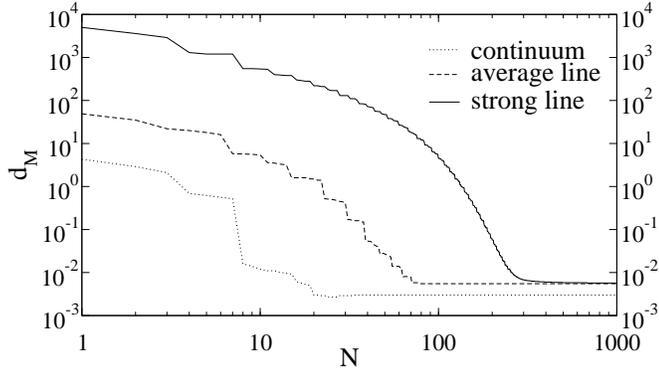}}
\caption{Evolution of the maximum relative error $d_\mathrm{M}$ with iteration
number $N$ for $n_\tau = 9$, $n_\mu = 5$ and $\tau_\mathrm{m} =
10^{-4}$.}
\label{fig3}
\end{figure}

The iterative algorithm was stopped after $N=1000$ iterations.
Figure \ref{fig3} shows that this number ensures convergence
of the ALI code, the convergence being slower when $\epsilon\to 0$ and
$\tau^* \to +\infty$.
Irregular steps in these curves are due to the Ng acceleration process, here operated every four iterations.

\subsection{The influence of $\tau_\mathrm{m}$}

We point out that $d_\mathrm{M}$ is improved when $\tau_\mathrm{m}$ goes to 0,
up to a given value where the accuracy is constant. 
Including the $\tau=0$ point in the grid and choosing $\tau_\mathrm{m} < 10^{-2}$, the best accuracy is warranted.
Excluding the $\tau=0$ point from the grid has no influence on accuracy if we choose $\tau_\mathrm{m} < 10^{-4}$.
The standard choice $\tau_\mathrm{m}=10^{-4}$ is thus correct, and this value will be adopted hereafter.

\subsection{The influence of $n_\tau$ and $n_\mu$}

\begin{figure}
\resizebox{\hsize}{!}{\includegraphics{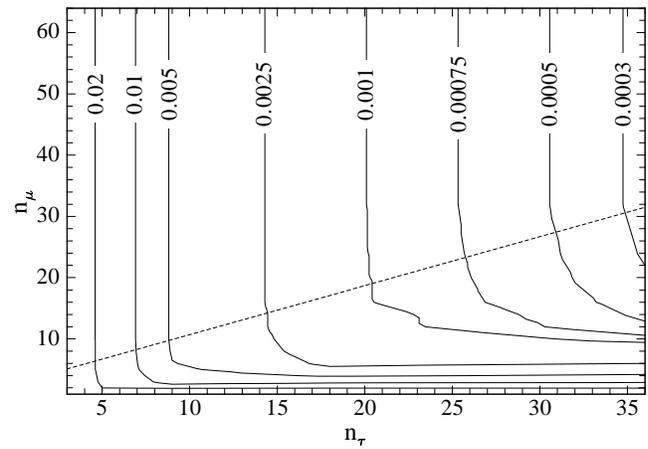}}
\caption{Maximum relative error $d_\mathrm{M}$ in an average line ($\tau^*=2000$,
$\epsilon = 10^{-4}$) as a function of $n_\tau$ and $n_\mu$.
The dashed curve represents a linear fit $n_\mu^\mathrm{(opt)} = 0.8 n_\tau + 2.7$.}
\label{fig4}
\end{figure}

In Fig.~\ref{fig4} are shown the variations of $d_\mathrm{M}$ in an average line as a function of $n_\tau$ and $n_\mu$.
The maximum relative error $d_\mathrm{M}$ decreases with increasing number $n_\tau$ of $\tau$-grid points, and it is sensitive to the choice of the number $n_\mu$ of angular
grid points up to an optimal value $n_\mu^\mathrm{(opt)}$; the
latter is defined as the smallest value of $n_\mu$ for which the
condition $|1-d_\mathrm{M}(n_\mu)/d_\mathrm{M}(64)| < 0.01$ holds.
The accuracy does not increase with a finer $\mu$-grid.
We note that $n_\mu^\mathrm{(opt)} < n_\tau$ and that we have a linear dependence of this optimal value on $n_\tau$: $n_\mu^\mathrm{(opt)} = 0.8 n_\tau + 2.7$ (dashed curve).
This fit is still valid for strong lines.

\begin{figure}
\resizebox{\hsize}{!}{\includegraphics{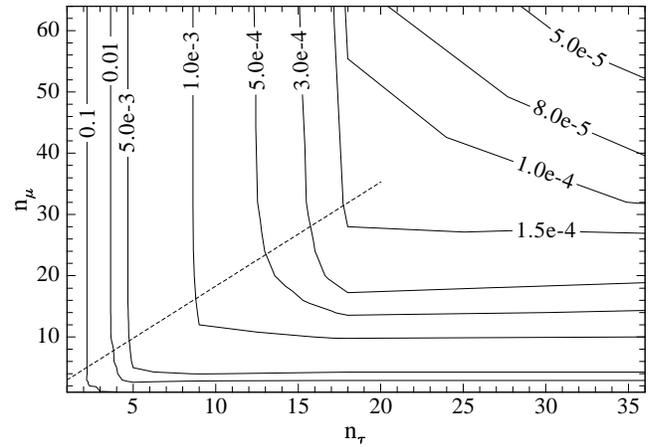}}
\caption{Maximum relative error $d_\mathrm{M}$ in the continuum ($\tau^*=20$,
$\epsilon = 0.01$) as a function of $n_\tau$ and $n_\mu$. Dashed curve
represents a linear fit $n_\mu^\mathrm{(opt)} = 1.7 n_\tau + 1.3$ for $n_\tau < 18$.}
\label{fig5}
\end{figure}

As seen in Fig.~\ref{fig5} (same as Fig.~\ref{fig4} for the continuum), the fit for lines cannot be applied to the continuum, for which $n_\mu^\mathrm{(opt)} > n_\tau$.
It is still possible to define and calculate an optimal value for $n_\tau < 18$, using the relation $n_\mu^\mathrm{(opt)} = 1.7 n_\tau + 1.3$ (dashed curve).
It appears that our ALI code is more demanding in angular resolution when solving the problem (\ref{eq_s2}) in a continuum than in a line.

\begin{figure}
\resizebox{\hsize}{!}{\includegraphics{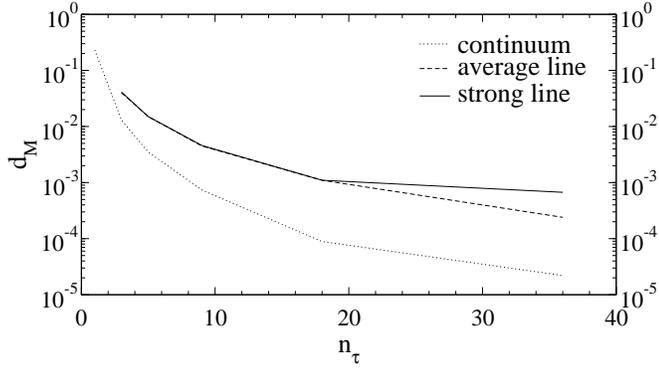}}
\caption{Maximum relative error $d_\mathrm{M}$ for different values of $n_\tau$ and $n_\mu=64$.}
\label{fig6}
\end{figure}

The results of Figs.~\ref{fig4} and \ref{fig5} are detailed in Fig.~\ref{fig6} for the three
chosen values of $(\epsilon,\tau^*)$ and $n_\mu = 64$.
We remark that the accuracy improves with $n_\tau$ for each couple
$(\epsilon,\tau^*)$, more significantly in the continuum than in lines.
\begin{figure}
\resizebox{\hsize}{!}{\includegraphics{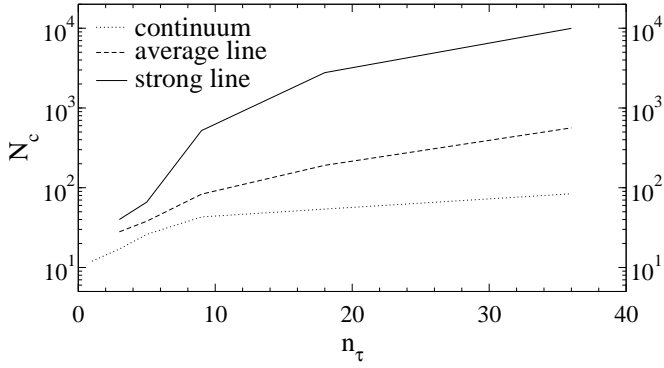}}
\caption{Number of iterations $N_\mathrm{c}$ used to reach convergence
($\varepsilon_\mathrm{c} = 10^{-2}$) for different values of the
parameter $n_\tau$ and $n_\mu=64$.}
\label{fig7}
\end{figure}
Figure~\ref{fig7} gives the number of iterations $N_\mathrm{c}$ used
to reach convergence for $\varepsilon_\mathrm{c} = 10^{-2}$ and $n_\mu
= 64$.
The number $N_\mathrm{c}$ appreciably increases with $n_\tau$ in the lines:
it is indeed well known that the rate of convergence of the one-point ALI
iterative scheme drops for an increasing refinement of the
spatial grid (Olson et al. 1986); however improvements were already
proposed (e.g., Trujillo Bueno \& Fabiani Bendicho 1995) in order
to increase significantly the rate of convergence of ALI-based methods.

\subsection{The influence of $\epsilon$ and $\tau^*$}

\begin{figure}
\resizebox{\hsize}{!}{\includegraphics{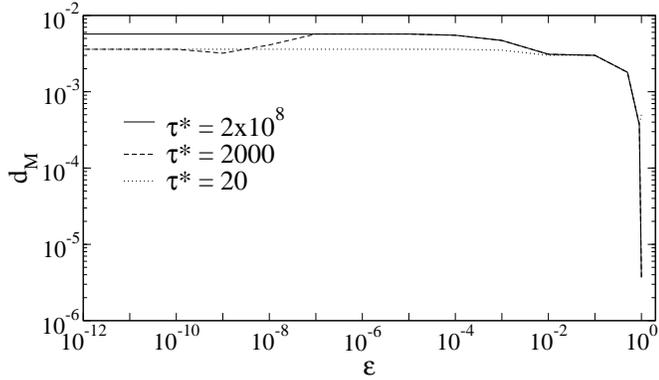}}
\caption{Maximum relative error $d_\mathrm{M}$ as a function of $\epsilon$ for
parameters $n_\tau = 9$ and $n_\mu = 5$.}
\label{fig8}
\end{figure}

\begin{figure}
\resizebox{\hsize}{!}{\includegraphics{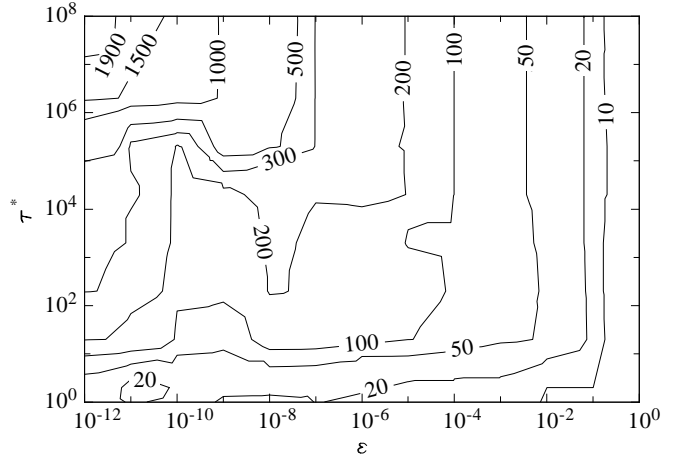}}
\caption{Number of iterations $N_\mathrm{c}$ ($\varepsilon_\mathrm{c}
= 10^{-2}$) as a function of $\epsilon$ and $\tau^*$ ($n_\tau = 9$,
$n_\mu = 5$).}
\label{fig9}
\end{figure}

The maximum relative error $d_\mathrm{M}$ and number of iterations $N_\mathrm{c}$ are shown 
in Figs.~\ref{fig8} and \ref{fig9} 
for an extended range of $(\epsilon,\tau^*)$ after the ALI code has
converged ($N=10\,000$, $n_\tau=9$ and $n_\mu=5$ are fixed here).

As seen in Fig.~\ref{fig8}, the accuracy hardly changes as
$\epsilon\to 0$ and $\tau^*\to +\infty$, but the number of iterations needed to achieve
convergence increases substantially (see Fig.~\ref{fig9}).
When $\epsilon>0.1$ the accuracy no longer depends on values of $\tau^*$.
The comparison of Figs.~\ref{fig6} and \ref{fig8} leads to a disagreement 
since the parameter $n_\mu$ is set to different values, 64 and 5 respectively.

In Fig.~\ref{fig9}, we have plotted the parameter $N_\mathrm{c}$ as a function of $\epsilon$ and $\tau^*$.
When $\epsilon\to 0$ and $\tau^* \to +\infty$, we note a slowing down of the convergence (already seen in Fig.~\ref{fig3}).

\section{Comments on previous studies}\label{sec_5}

Now we compare our results for the monochromatic scattering problem with those published by Trujillo Bueno \& Fabiani Bendicho (1995) and Trujillo Bueno \& Manso Sainz (1999). Although these two papers concern mainly the development of new iterative methods for radiative transfer applications (for the unpolarized and polarized cases respectively) they give some information on the accuracy of the numerical solutions obtained for spatial grids of increasing resolution.

In Table~\ref{tab1}, good agreement is found between our values of
$N_\mathrm{c}$, $d(\epsilon, \tau^*, 0)$ and those given by these authors; our surface
relative error $d(\epsilon, \tau^*, 0)$ corresponds to their surface
true error $T_\mathrm{e}$.
The observed small discrepancies are possibly due to the different scattering
laws adopted, leading to different $\Lambda$-operators.
\begin{table*}
\caption{Comparison of results obtained with our (ALI+Ng) code and
previous ones.
Our $N_\mathrm{c}$ is defined by $\varepsilon_\mathrm{c} = 0.01$, while values from other authors are
based on a graphical guess $\varepsilon_\mathrm{c} \approx 0.05$.
The optical thickness is $\tau^* = 2\times 10^8$.
Note that in Trujillo Bueno \& Manso Sainz (1999), $N_\mathrm{c}$ values (in parenthesis) are given for a
non-accelerated Jacobi scheme.
These numbers have been divided by 2 in order to estimate the number of iterations when Ng acceleration is used.}
\label{tab1}
\centering
\begin{tabular}{llllllll}
\hline
\noalign{\smallskip}
$\epsilon,n_\tau,n_\mu$ & \multicolumn{2}{l}{JTB \& PFB (1995)} & \multicolumn{2}{l}{JTB \& RMS (1999)} & \multicolumn{3}{l}{This article} \\
\noalign{\smallskip}
\hline
\noalign{\smallskip}
& $N_\mathrm{c}$ & $T_\mathrm{e}$ & $N_\mathrm{c}$ & surface $T_\mathrm{e}$ & $N_\mathrm{c}$ & $d_\mathrm{M}(\epsilon,\tau^*)$ & $d(\epsilon,\tau^*,0)$  \\
\noalign{\smallskip}
\hline\noalign{\smallskip}
$10^{-6}$, 9, 1  & 180 & $3.5\times 10^{-3}$ && & 179 & $8.6\times 10^{-2}$ & $4.1\times 10^{-3}$ \\ 
$10^{-12}$, 9, 1 & 1300 & $3.5\times 10^{-3}$ &&& 985 & $8.6\times 10^{-2}$ & $4.1\times 10^{-3}$ \\ 
$10^{-4}$, 5, 64 && & 33 (65) & $2\times 10^{-2}$ & 42 & $1.5\times 10^{-2}$ & $1.3\times 10^{-2}$\\ 
$10^{-4}$, 9, 64 && & 75 (150) & $3\times 10^{-3}$ & 88 & $4.5\times 10^{-3}$ & $3.9\times 10^{-3}$\\ 
$10^{-4}$, 18, 64 && & 175 (350) & $4\times 10^{-4}$ & 184 & $1.1\times 10^{-3}$ & $9.0\times 10^{-4}$ \\ 
$10^{-4}$, 36, 64 && &  400 (800) & $5\times 10^{-5}$ & 356 & $2.4\times 10^{-4}$ & $1.9\times 10^{-4}$ \\ 
\hline
\end{tabular}
\end{table*}
\begin{figure}[ht]
\resizebox{\hsize}{!}{\includegraphics{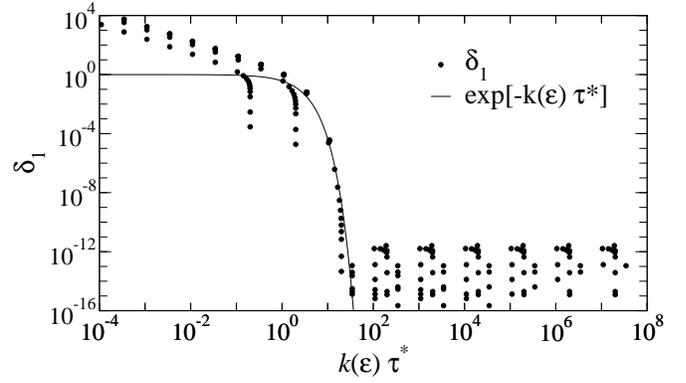}}
\caption{Relative difference
$\delta_1=|1-S(\epsilon,\tau^*,0)/\sqrt{\epsilon}|$ as a function of $k(\epsilon)\tau^*$.
Dots represent values of $\epsilon \in [10^{-12},1]$ and $\tau^* \in [0.2,2\times
10^8]$, and the solid curve is $\exp\left(-k(\epsilon)\tau^*\right)$.}
\label{fig10}
\end{figure}

In Fig.~\ref{fig10} it is shown how far the semi-infinite exact result
$S(\epsilon,0)=\sqrt{\epsilon}$ agrees with the finite one.
This comparison is useful since many authors use the $\sqrt{\epsilon}$-law
as a check for their calculations in thick slabs.
We plot the relative difference
$\delta_1=|\,1-S(\epsilon,\tau^*,0)/\sqrt{\epsilon}\,|$ as a function of
$k(\epsilon) \tau^*$, and the quantity
$\exp\left(-k(\epsilon)\tau^*\right)$ which characterizes the
validity of the $\sqrt\epsilon$-law (solid curve).  For $k(\epsilon)\tau^* > 100$,
the accuracy limit of our code is reached, which explains that the
solid curve no longer fits the dots.  This law is very well
satisfied in lines ($k(\epsilon)\tau^* \approx 34$ in an average line)
but not enough in the continuum ($k(\epsilon)\tau^* \approx 3.3$).
We conclude that the $\sqrt\epsilon$-law can be used as a test for the
ALI code when $k(\epsilon)\tau^* > 10$, since then $\sqrt\epsilon$ is an
approximation to the surface value with an accuracy better than $10^{-4}$, as seen in Fig.~\ref{fig10}.

In Trujillo Bueno \& Fabiani Bendicho (1995), the Eddington approximation is used as the reference solution for a one-point angular quadrature $n_\mu=1$ with $\mu=\pm 1/\sqrt 3$.
The analytical expression of the Eddington approximation in a finite slab is:
\begin{eqnarray}
\label{eq_sedd}
\lefteqn{S_\mathrm{E}(\epsilon,\tau^*,\tau) = 1 - (1-\epsilon) \frac{\exp\left(-\sqrt{3\epsilon}\,\tau\right) + \exp\left(-\sqrt{3\epsilon}\,(\tau^* - \tau)\right)}{1+\sqrt\epsilon+(1-\sqrt\epsilon)\exp\left(-\sqrt{3\epsilon}\,\tau^*\right)}. } \nonumber \\
\end{eqnarray}

\begin{figure}[ht]
\resizebox{\hsize}{!}{\includegraphics{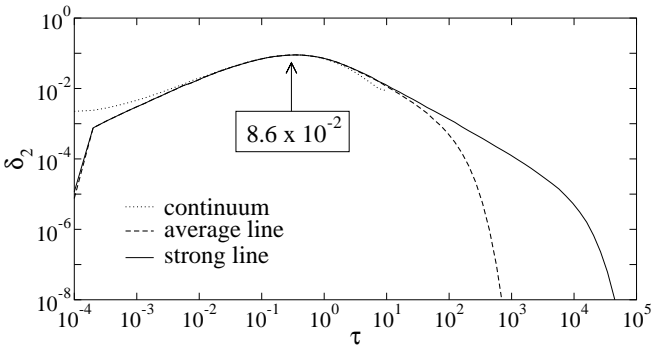}}
\caption{Variation with $\tau$ of the function $\delta_2(\epsilon,\tau^*,\tau) = |1 - S_\mathrm{E}(\epsilon,\tau^*,\tau)/S(\epsilon,\tau^*,\tau)|$ characterizing the accuracy of  the Eddington approximation $S_\mathrm{E}$ as given by (\ref{eq_sedd}) for the usual three couples $(\epsilon,\tau^*)$.}
\label{fig11}
\end{figure}

This is the exact solution of the monochromatic scattering problem when the mean intensity is calculated with the above-mentioned one-point angular quadrature.
However, as is well-known, it gives only an approximation to the exact (i.e., multi-angle) solution of the full problem (\ref{eq_s1})-(\ref{eq_e1}).
In other words, the true error given by Trujillo Bueno \& Fabiani Bendicho (1995) is relative to the $n_\mu=1$ monochromatic scattering problem only, it does not give information on the error that would have been got by comparing the numerical solution to the $n_\mu=1,3,5, \ldots$ problem to the exact multi-angle solution ($n_\mu=\infty$).
In fact, as given in Table~\ref{tab1} for $n_\mu=1$, when the solution of the $n_\mu=1$ problem is compared to the exact multi-angle solution, we find that the maximum error for $n_\tau=9$ is $8.6\times 10^{-2}$.
The latter represents the maximum relative difference between the Eddington approximation and the exact solution (Fig.~\ref{fig11}).
A similar investigation, but for the two-level atom resonance-line scattering polarization problem, was carried out by Trujillo Bueno \& Manso Sainz (1999), whose Table 3 gives the surface true-error values of the fractional atomic polarization for $n_\mu=3,5,7,11, \ldots, 61$.

\section{Conclusion}

Our ALI code has been subjected to a wide range of tests, revealing at
the same time its capabilities and its limits.  Before developing
these two points, we note that our conclusions are relative to the
particular code we have used (based on Jacobi's method), specifically solving the
standard problem (\ref{eq_s1})-(\ref{eq_e1}).
The accuracy of the code is ultimately determined by the accuracy of the
formal solver we have used (parabolic short characteristics).

We have checked the great robustness of our code, which is certainly
its most remarkable feature.  It is able to solve the
standard problem (\ref{eq_s1})-(\ref{eq_e1}) for a wide range of input
parameters $\epsilon$ and $\tau^*$, with no important lack of
performance when $\epsilon \to 0$ and/or $\tau^* \to
+\infty$.

However, the lowest accuracy of the ALI numerical solutions happens in
the outermost layers of a star, corresponding to $\tau$ lower than the
thermalization depth $1/k(\epsilon)$, these layers forming, by
definition, the atmosphere of the star.
The accuracy of our code is not better than, say $10^{-2}$, when we choose 
$n_\tau = 9$, $n_\mu = 5$ and limit the number of iterations to $N<100$, as it is currently done in stellar atmospheres modelling.
To improve the accuracy of the calculations up to $\sim 10^{-3}$, the parameters $n_\tau$, $n_\mu$ and $N$ should be
set to larger (but today impractical) values when solving the radiative transfer equation on a large frequency spectrum, i.e. at thousands of frequencies.
Indeed we pointed out a truly noticeable improvement of
the accuracy when using finer grids in $\tau$ or $\mu$.
Such an observation was made easier by the use of a very accurate
reference solution.
Of course, increasing the level of refinement of
both spatial and angular quadratures has a strong impact upon
the number of iterations needed for convergence.
However, to overcome this difficulty while keeping the same accuracy on the numerical
solutions, methods based on Gauss-Seidel and successive
over-relaxation iterations were already proposed by Trujillo
Bueno \& Fabiani Bendicho (1995).

Another important question is relative to the propagation of errors in
a stellar atmosphere model: to what extent are the main 
quantities provided by the model (populations of heavy particles,
electron density, pressure, etc.) sensitive to the accuracy on the
RTE solution? We intend to tackle this subject in a future work by
constructing an accurate -- but still very idealized -- stellar
atmosphere model, in which the main quantities are first derived from
an exact solution to the RTE, and then from the solution given by a ALI-based numerical method.

\begin{acknowledgements}
The authors wish to thank M. Ahues, A. Largillier, G. Panasenko (Numerical Analysis team of the
University Jean Monnet of Saint-Etienne, France), A. Amosov (Moscow Power Engineering Institute,
Russia) and J. Bergeat (Centre de Recherche Astronomique de Lyon) for some helpful discussions concerning this work.
We also thank Ivan Hubeny and Javier Trujillo Bueno for their valuable comments on a previous version of our manuscript.
\end{acknowledgements}

\end{document}